\documentclass[preprint2]{aastex}
\usepackage{natbib}

\newcommand{\eg}{e.g., }
\newcommand{\ie}{i.e., }
\newcommand{\Msun}{M_{\odot}}
\newcommand{\kms}{km~s$^{-1}$}
\newcommand{\ergs}{erg~s$^{-1}$}

\def\gsim{\mathrel{\rlap{\lower 4pt \hbox{\hskip 1pt $\sim$}}\raise 1pt
\hbox {$>$}}}
\def\lsim{\mathrel{\rlap{\lower 4pt \hbox{\hskip 1pt $\sim$}}\raise 1pt
\hbox {$<$}}}

\shorttitle{3D Models for HVFs in SNe Ia}
\shortauthors{Tanaka et al.}
\slugcomment{Accepted for publication in the Astrophysical Journal}
\begin{document}

\title{3D Models for High Velocity Features in Type Ia Supernovae }
\author{
Masaomi Tanaka\altaffilmark{1},
Paolo A. Mazzali\altaffilmark{1,2,3},
Keiichi Maeda\altaffilmark{4}, and
Ken'ichi Nomoto\altaffilmark{1}
}

\altaffiltext{1}{Department of Astronomy, Graduate School of Science, University of Tokyo, Hongo 7-3-1, Bunkyo-ku, Tokyo 113-0033, Japan; mtanaka@astron.s.u-tokyo.ac.jp, nomoto@astron.s.u-tokyo.ac.jp}
\altaffiltext{2}{Max-Planck-Institute f$\ddot{u}$r Astrophysik, Karl-Schwarzschild-Str., 1, D-85741 Garching bei M$\ddot{u}$nchen, Germany;mazzali@MPA-Garching.MPG.DE}
\altaffiltext{3}{National Institute for Astrophysics-OATs, Via Tiepolo 11, I-34131 Trieste, Italy}
\altaffiltext{4}{Department of Earth Science and Astronomy,
Graduate School of Arts and Science, University of Tokyo, Meguro-ku, Tokyo
153-8902, Japan; maeda@esa.c.u-tokyo.ac.jp}

\begin{abstract}
Spectral synthesis in 3-dimensional (3D) space for the earliest spectra of 
Type Ia supernovae (SNe Ia) is presented. In particular, the high velocity
absorption features that are commonly seen at the earliest epochs ($\sim 10$
days before maximum light) are investigated by means of a 3D Monte Carlo
spectral synthesis code. The increasing number of early spectra available
allows statistical study of the geometry of the ejecta. The
observed diversity in strength of the high velocity features (HVFs)  can be
explained in terms of a ``covering factor'', which represents the fraction of the projected photosphere that is concealed by high velocity material. Various geometrical
models involving high velocity material with a clumpy structure or a thick
torus can naturally account for the observed statistics of HVFs. HVFs may be
formed by a combination of density and abundance enhancements. Such
enhancements may be produced in the explosion itself or may be the result of
interaction with circumstellar material  or an accretion disk. Models with 1 or
2 blobs, as well as a thin torus or disk-like enhancement are unlikely
as a standard situation.
\end{abstract}


\section{Introduction}
\label{sec:intro}

Type Ia supernovae (SNe Ia) have found use in cosmology  after it has been
established that the maximum luminosity can be calibrated 
via an empirical relation
between it and the shape of the light curve (LC, Phillips 1993; Riess, Press
\& Kirshner 1996). The origin of this empirical relation is however not fully understood
(\eg Mazzali et al. 2001),
mainly because of uncertainties regarding the properties of the explosion
mechanism.

A one-parameter ordering scheme similar to the one found for the LC was also
proposed for the spectra (Nugent et al. 1995). In this scheme temperature is
the driving parameter characterizing the spectra. While this relation may apply
near maximum, detailed observations of SNe Ia with extended early time spectral
coverage indicate that SNe Ia with similar LCs may have different absorption
line velocities. In particular, the absorptions of Si II $\lambda$ 6355, S II
$\lambda$ 5640, Ca II H \& K and Ca II IR triplet have been investigated by
several authors (\eg Patat et al. 1996; Hatano et al. 2000; Kotak et al. 2005;
Benetti et al. 2005).

Simultaneously, the number of polarization observations has increased 
(SN 1996X, Wang et al. 1997; SN 1999by, Howell et al. 2001; 
SN 2001el, Wang et al. 2003; SN 2004dt, Wang et al. 2004) 
and it has become possible to infer asphericity by means of spectropolarimetry.
These results suggest that some SNe Ia have a
somewhat aspherical geometry of the photosphere
($\gsim 10$\% for SN 2001el or $\sim 20$\% for SN 1999by, 
assuming a ellipsoidal geometry). Even larger degree of asphericity 
is inferred for the distribution of intermediate mass elements 
(at the level of $\lsim 25$ \%; see Wang et al. 2003). 
Multi-dimensional numerical simulations have also suggested 
that the explosion is aspherical, mainly owing
to the nature of the deflagration flame in 3D numerical simulations (Reinecke,
Hillebrandt \& Niemeyer 2002; Gamezo et al. 2003; R\"opke \& Hillebrandt 2005).

Recently, high velocity features (HVFs) in the Ca II IR triplet have been the
subject of interest: SN 1994D (Hatano et al. 1999), SN 1999ee (Mazzali et al.
2005a), SN 2000cx (Thomas et al. 2004), SN 2001el (Wang et al. 2003), SN 2003du
(Gerardy et al. 2004), SN2005cg (Quimby et al. 2006).  Here, HVFs are defined
as absorptions with velocities much higher than the photospheric component,
which has a typical velocity $v \sim$ 15,000 \kms \ at $\sim 1$ week before
maximum. HVFs are often seen detached from the photospheric component, at
velocities of 17,000--29,000 \kms. The above papers discuss  the possible
origin of HVFs including primordial metallicity (Lentz et al. 2000), a property
of the explosion and interaction of the SN ejecta with circumstellar matter
(CSM). Whatever their origin, understanding HVFs can cast light on the
explosion mechanism.

Mazzali et al. (2005a) tried to fit the HVFs of SN 1999ee using abundance or
density enhancements. Their results suggest that it is impossible to reproduce
HVFs by abundance enhancement only and $\sim 0.1\Msun$ of additional material
is required if only the density enhancement is assumed. 
Gerardy et al. (2004) and Quimby et al. (2006) discussed 
how HVFs may result from interaction with a CSM.
They obtained good agreement for several features including Mg II, 
and found $2 \times 10^{-2} \Msun$ and $ 5-7 \times 10^{-3} \Msun$ 
of the CSM are needed for SN 2003du and SN 2005cg, respectively.
It should be noted that the explosion model (\ie abundance distribution and 
the density structure) and the way to introduce additional material 
are not identical with each other. Recently, Mazzali et al.
(2005b) have shown that almost all SNe Ia have Ca II IR triplet HVFs at the
earliest epochs.  These HVFs have velocities ranging from 17,000 \kms\ to
29,000 \kms\ in different supernovae. The ubiquity of HVFs may indicate that
they do not come from an extreme environment. Probably, the combination of
different phenomena results in the observed HVFs.

The geometry of the ejecta has been studied in multi-dimensional space. Kasen
et al. (2003) analyzed both spectroscopy and spectropolarimetry of SN 2001el
and showed that both an aspherical photosphere and a single high velocity blob
can reproduce the observations. Thomas et al. (2004) derived a similar geometry 
for the Ca II HVFs of SN 2000cx. 

Although all previous studies performed modeling for each SN, no systematic
study has been made of the 3D properties of the geometry. We present here
synthetic spectra computed in 3D space for various geometries and show how HVFs
are affected by different geometrical configurations and line-of-sight
effects.  The increasing number of early spectra available enables us to
investigate the statistical properties of HVFs, and to constrain the geometry
of the ejecta and possibly the nature of the explosion. 

In \S \ref{sec:HVF1D}
we discuss how HVFs are formed and suggest a possible parametrized
description.
In \S \ref{sec:method} 
we present the method of calculation for the spectra and our models.
In \S \ref{sec:results} 
the results and some properties of 3D computations are presented.
In \S \ref{sec:dis} 
the probable geometries that can reproduce the observed trend are discussed. 
Finally, conclusions including considerations on the origin of the high 
velocity material and on the asphericity of the explosion are made 
in \S \ref{sec:con}.

\section{Properties of the HVFs}
\label{sec:HVF1D}

HVFs come in different forms. Figure \ref{fig:HVF} shows the CaII IR triplet in
the earliest spectra of SN 1999ee ($-9$ days before B maximum), SN 2002dj
($-11$ days), SN 2003cg ($-8$ days), and SN 2003du ($-11$ days). At that epoch,
the photospheric velocity is $v_{\rm ph} \sim 12,000 - 14,000$\,\kms. The
photospheric component of the CaII IR triplet has $v \sim 14,000 -
16,000$\,\kms\ in all SNe Ia, as suggested by the data of Mazzali et al.
(2005b).  This is slightly higher than the photospheric velocity, since the
CaII IR triplet is stronger than SiII or SII, and forms above the photosphere. 
However, the velocities of the absorption minima in SN 1999ee ($\sim 20,000$ km
s$^{-1}$), SN 2002dj ($\sim 25,000$ km s$^{-1}$) and SN 2003du ($\sim 20,000$
km s$^{-1}$) are much higher, suggesting that these lines form well above the
photosphere. Therefore the HVFs are expected to have a different origin that
the photospheric absorption.  Comparing SN 2002dj and SN 2003du, which were
observed at the same epoch, it can be noticed that their HVFs have different
strength and velocity range.  The same is true for SN 1999ee and SN 2003du,
although their spectra were not taken at exactly the same epoch.  At very early
epochs, even a small difference in epoch can strongly affect the strength of
the photospheric component. However, judging from a number of early spectra we
can safely conclude that there are real variations in the strength of the
HVFs.  Therefore we examine a parameterization of the properties of the HVFs
based on their velocity range ($v_{\rm HV}$), their strength ($S_{\rm HV}$) and
on the strength of the photospheric component ($S_{\rm ph}$). 

We performed numerical tests to verify what governs the properties of the HVFs.
The Monte Carlo SN spectrum synthesis code described in Mazzali \& Lucy (1993),
Lucy (1999) and Mazzali (2000) was used. The code computes a synthetic spectrum
based on the luminosity, the position of the photosphere, the time since the
explosion, the density structure and the abundance distribution. The density
structure of the standard deflagration model W7 (Nomoto, Thieleman \& Yokoi
1984) was used in all the calculations presented in this section. Photospheric
velocity ($v_{\rm ph} = 12,500$\kms), epoch (8 days since the explosion 
assuming a
risetime of 19 days) and luminosity ($L=3.0 \times 10^{42}$\ergs) were fixed
for simplicity.

First we consider the case where $S_{\rm ph}$ and $v_{\rm HV}$ are fixed and
parameterize $S_{\rm HV}$ through the line optical depth  of the CaII IR
triplet. We assume that the high velocity region lies at  $v_{\rm HV} \sim
22,000 - 29,000$ \kms. In this region, the line optical depth is imposed
irrespective of the value computed  consistently with the W7 structure in order to
simplify the investigation of the behavior of HVFs. For example, 
if line optical
depth is taken as $1 - \exp (- \tau) = 0.5$, about
half the photons that come into resonance with the CaII IR triplet are
scattered. Figure \ref{fig:SHV1D} shows HVFs with various line optical depths
$1 - \exp (-\tau) = 0.1 - 0.9$. With this approximation the depth of the HVFs
depends almost linearly on $1 - \exp (-\tau)$. 
The origin of the observed diversity in
strength is investigated in 3-dimensional space in \S \ref{sec:method} and \S
\ref{sec:results}.

Next we show how the strength of the photospheric component affects the 
absorption feature. Here $S_{\rm ph}$ is parameterized through the abundance of
Ca. In the calculation, a homogeneous abundance distribution is used, which is
a reasonable assumption because the region used in the computation is only $v
\gsim 12,500$\kms.  Figure \ref{fig:Sph1D} shows the CaII IR triplet with
various strengths of the photospheric component. The mass fraction of Ca is
$X({\rm Ca}) = 0.068, 0.034, 0.017, 0.008$, and 0.04, respectively. The
photospheric  absorption depth does not change linear with the abundance, and
the line becomes saturated if the abundance is more than $\sim 1$\%. Since in
this calculation the line optical depth at high velocity is fixed as $1 - \exp
(-\tau) = 0.7$ and the velocity separation is large, the HVFs are not affected
by the photospheric component.

Using only the abundance of Ca to change the strength of the photospheric
component is not completely realistic because the line strength also depends
strongly on temperature and ionization state.  Ca is mostly CaIII at the
typical temperature of SNe Ia before maximum ($T \sim 12,000$K) and only a
small fraction is CaII and CaIV.
The strength of the CaII IR triplet is
proportional to temperature because of the behavior of the ionization.  We
tested models with different luminosities, which yield different temperature
structures. We found that even with the highest Ca abundance 
$(X({\rm Ca})=0.068)$, when  the luminosity is increased 
by a factor of $\sim 4$, the strength of the CaII IR triplet becomes 
weaker than in the original model 
with the lowest Ca abundance, $X({\rm Ca}) = 0.002$.  An inverse correlation
between the strength of the photospheric component and the luminosity at these
early epochs might therefore be expected.  To verify this, more observations
are needed and it is necessary to fit these individually to determine the exact
photospheric velocity and temperature. This is beyond the scope of this work,
and will be the subject of a separate investigation. (M. Tanaka et al., in
preparation). 

\begin{figure*}
\begin{tabular}{cc}
\includegraphics[scale=0.80]{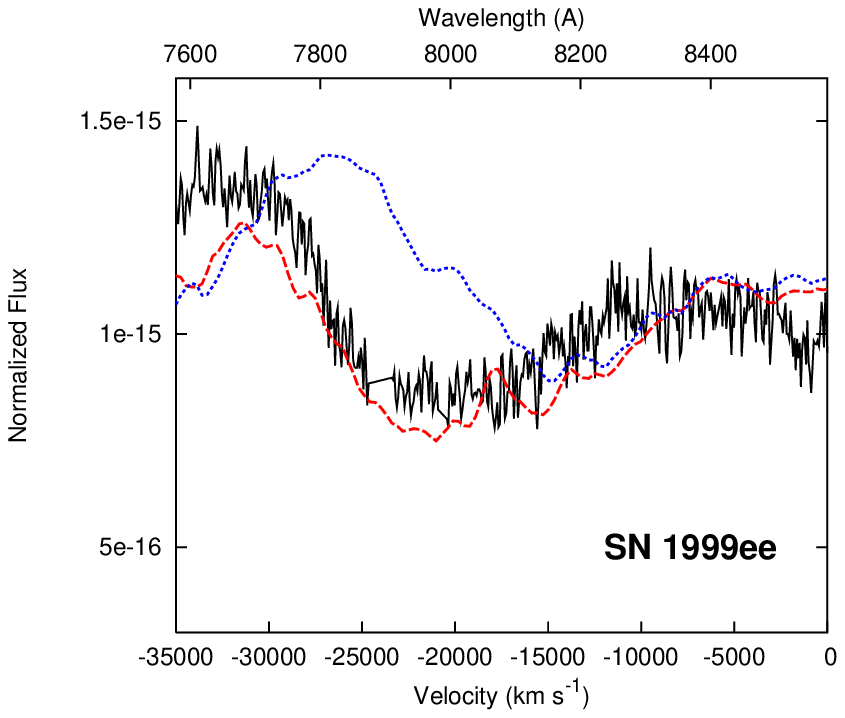} &
\includegraphics[scale=0.80]{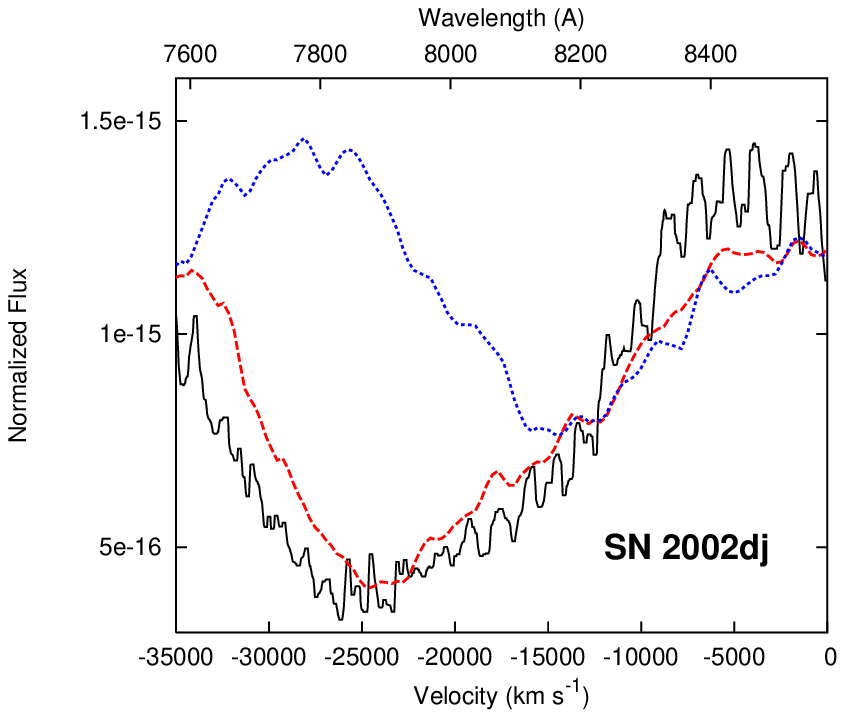} \\
\includegraphics[scale=0.80]{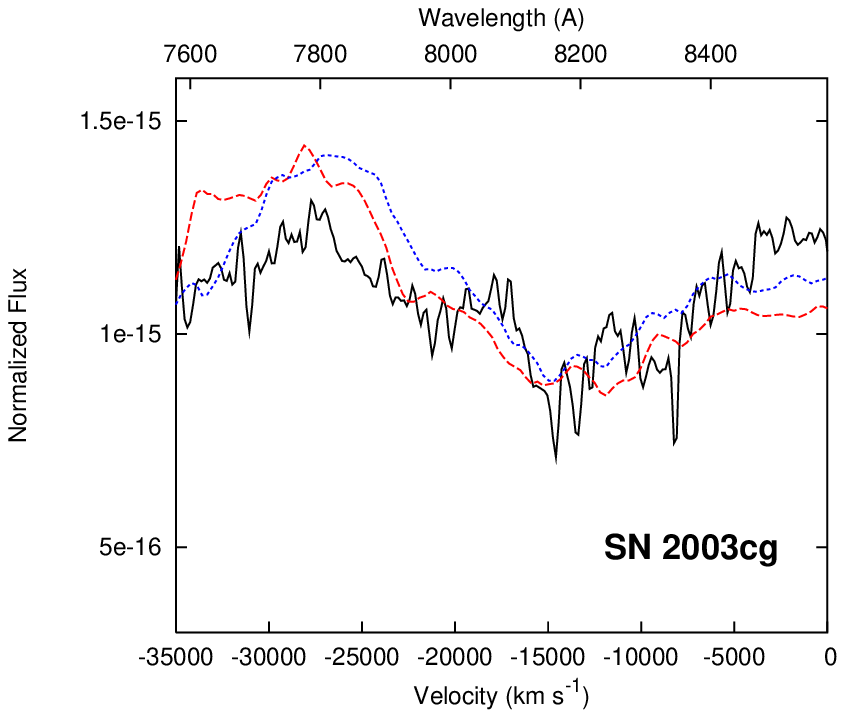} &
\includegraphics[scale=0.80]{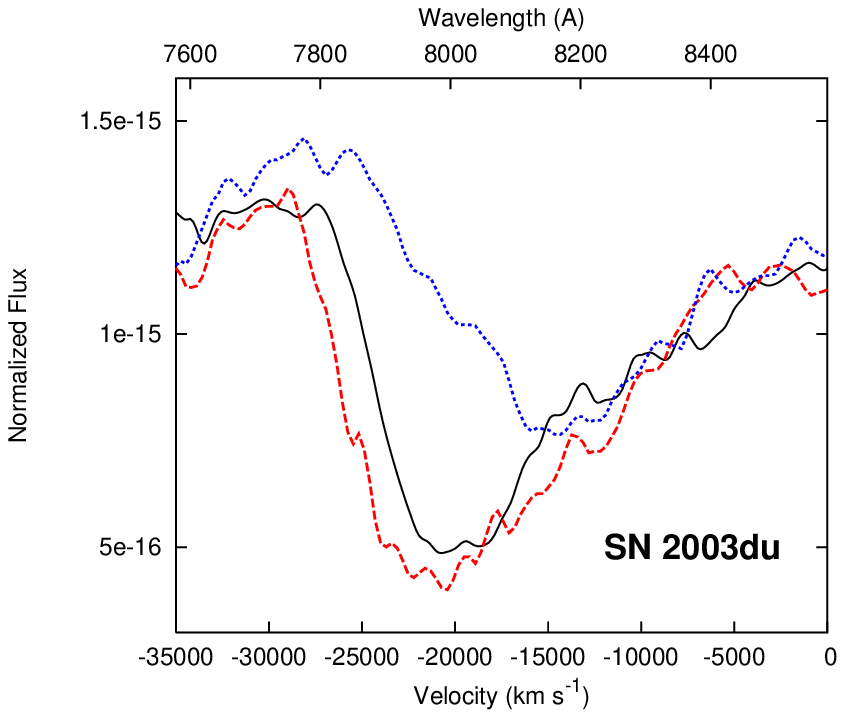} \\
\end{tabular}
\caption{The observed CaII IR triplet (solid lines) compared with synthetic 
spectra computed by adjusting all three parameters
($S_{\rm HV}, v_{\rm HV}$ and $S_{\rm ph}$.
All spectra are normalized to their continuum level.
The red dashed lines are the line profiles computed including the high velocity 
components, while the blue dotted lines
are the profiles computed with only the photospheric component.
$S_{\rm HV}$ corresponds to $1 - \exp (-\tau)$, where $\tau$ is the 
line optical depth of the CaII IR triplet in the high velocity region.
$S_{\rm ph}$ is varied through the abundance of Ca [$X({\rm Ca})$].
SN 1999ee (9 days before maximum, -9days; top left) 
is reproduced with $1-\exp (-\tau) = 0.5$, 
$v_{\rm HV} \sim 22,000 - 27,000$\kms, and $X({\rm Ca}) = 0.004$.
SN 2002dj (-11 days; top right) is reproduced with $1-\exp (-\tau) = 0.7$, 
$v_{\rm HV} \sim 22,000 - 29,000$\kms, and $X({\rm Ca}) = 0.008$.
SN 2003cg (-8 days; bottom left) is reproduced with $1-\exp (-\tau) = 0.2$, 
$v_{\rm HV} \sim 22,000 - 25,000$\kms, and $X({\rm Ca}) = 0.004$.
SN 2003du (-11 days; bottom right) is reproduced with $1-\exp (-\tau) = 0.9$, 
$v_{\rm HV} \sim 22,000 - 25,000$\kms, and $X({\rm Ca}) = 0.008$.
\label{fig:HVFfit}}
\end{figure*}

The results of the tests we performed by changing $S_{\rm HV}$ and $S_{\rm ph}$
(Figures \ref{fig:SHV1D} and \ref{fig:Sph1D}) show clearly that changing these
two parameters is not enough to reproduce all the observed HVFs. 
As seen between SN 2002dj and SN 2003du in Figure \ref{fig:HVF} 
(and see Figure 4 in Quimby et al. 2006),
the width of the high velocity absorption is not identical among SNe that
have a similar depth of the HVFs.
No combination of $S_{\rm HV}$ and $S_{\rm ph}$, however, can account for 
such a difference.
Therefore we conclude that the additional
parameter $v_{\rm HV}$ is also needed. One might think that this is because of
our assumption of spherical symmetry. However, we obtain the same results in 3D
geometry (\S \ref{sec:results}). Figure \ref{fig:HVFfit} shows how the three
parameters ($S_{\rm HV}, v_{\rm HV}$ and $S_{\rm ph}$) work. Their combination
determines the depth and shape of the absorption and the position of 
the absorption minimum. 
Although the luminosity and the photospheric velocity are not
fine-tuned, the shape of the absorption feature is reproduced well.

The numerical tests presented here may be somewhat unrealistic because we
assumed a detached spherical shell that is optically thick to the the CaII line
and it is probably more realistic to consider 3D structures. The explosion may
be aspherical and has fragmentations (yielding a clumpy structure), or
interaction between the SN ejecta and a circumstellar disk may occur (yielding
a torus-like structure). In the next section, we use 3D models to see what the
three parameters used in this section (i.e., $S_{\rm HV}, v_{\rm HV}$, and
$S_{\rm ph}$) physically mean.

\section{Method and Models in 3D space}
\label{sec:method}
\subsection{3D Monte Carlo Code}
\label{sec:code}

To investigate 3D effects on the HVFs, synthetic spectra are calculated 
assuming various geometries (\S \ref{sec:model}). We have developed a
3-dimensional Monte Carlo radiative transfer code based on the 1D Monte Carlo
code described in Mazzali \& Lucy (1993), Lucy (1999), and Mazzali (2000). The
code assumes the Sobolev approximation and a sharply defined photosphere above
which there is no energy deposition. Spherical coordinates are used in the code
in order to treat the photosphere precisely. The number of meshes depends on
the model and is typically $20 \times 20 \times 20$ for $r, \cos(\theta)$ and
$\phi$, respectively.  
The path of the energy packets is traced in 3D space, and
the effect of back scattering into the photosphere is included. Line treatment
is free from the approximation of resonance scattering, \ie photon branching
is considered correctly as outlined by Lucy (1999). The free parameters are the
luminosity $L(\theta, \phi)$,  the photospheric velocity $v_{\rm ph}(\theta,
\phi)$  and the epoch since the explosion $t$, all of which are defined as in
the 1D code. Though angle dependent $L$ and $v_{\rm ph}$ can be used in the
code, here we assume them to be uniform for simplicity. We again assume $v_{ph}
= 12,500$\kms, $t=8.0$ days since the explosion and $L=3.0 \times 10^{42}$\ergs as
typical values for the earliest spectra of SNe Ia. With these
parameters, the temperature structure, excitation and ionization are computed
in all zones using a modified nebular approximation (Mazzali \& Lucy 1993).
This assumes that there is no net exchange of energy  between matter and
radiation.  Photon flux is collected by recording energy packets at the outer
boundaries of the ejecta and is binned into a $50 \times 50$ solid angle mesh.
Since the path of each packet is affected by the density and temperature
structure, the emergent spectrum depends on the orientation.

\subsection{Models}
\label{sec:model}

Our models are based on the spherical deflagration model W7 
(Nomoto et al. 1984). We introduced additional material in the outer
layers of the ejecta to produce HVFs. We are not concerned here with the origin
of this material, which may come from fluctuations of the explosion or from
interaction with CSM or an accretion disk.
Abundance enhancements are not considered because they
have been shown not to be suitable for reproducing HVFs (Mazzali et al. 2005a).
The velocity range and the degree of density enhancement are determined by
fitting one of the strongest CaII HVF, that of SN 2002dj, with the 1D code. As
a result, $\sim 0.1 \Msun$ of material at $v \sim 22,000-29,000$\kms\ is added
in a shell.

We fixed this degree of density enhancement in our 3D computations. We used a
density enhancement rather than an artificial line strength, which was used in
the previous section, because we want to explore the cause of HVFs. In fact, a
spherical density enhancement has the same effect on the CaII IR triplet
profile as the artificial line strength. However, other lines, such as SiII
$\lambda$6355, are also affected when using a density enhancement. If the
observed behavior of both the CaII IR triplet and SiII $\lambda$6355 is
reproduced consistently with a density enhancement, we may conclude that HVFs
are due to density enhancements. Additionally, the mass invoked in HVFs  should
be reduced in 3D space. 
In Mazzali et al. (2005a), a large amount of material ($\sim 0.10 \Msun$
in the case of density enhancement only) was
required to produce HVFs, and this was a serious concern in the hypothesis that
HVFs come from either the explosion or CSM interaction.
Although Gerardy et al. (2004) and Quimby et al. (2006) found 
only $\sim 2 \times 10^{-2} \Msun$ and $\sim 5-7 \times 10^{-3} \Msun$ 
are enough for the HVF of SN 2003du and SN 2005cg, respectively,
the HVF that has a higher velocity and a deeper absorption has been observed 
like SN 2002dj (see Fig. \ref{fig:HVF}).
We will give an estimate of the amount of material needed in a 3D calculation.

We mapped the density enhancement into 3D space conserving the radial velocity range and the
degree of enhancement derived in the 1D calculations. 
We tested various
morphologies that may be realistic, including one or two large blobs, a small
number of discrete blobs, a crowded clumpy structure, and tori of various
opening angles.

\section{Results}
\label{sec:results}

\begin{figure*}
\begin{tabular}{cc}
\includegraphics[scale=.80]{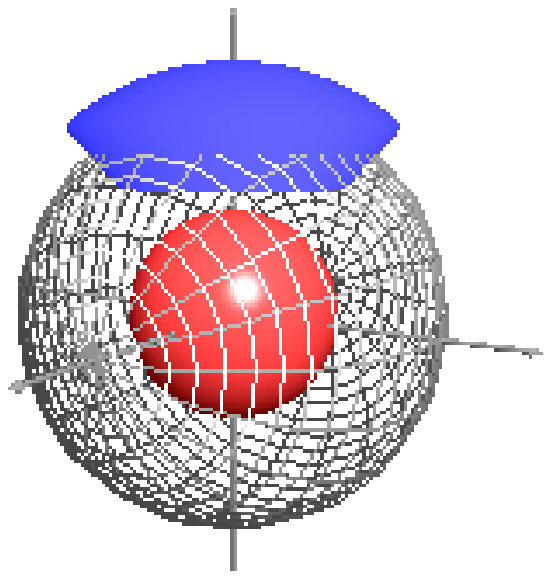} &
\includegraphics[scale=.90]{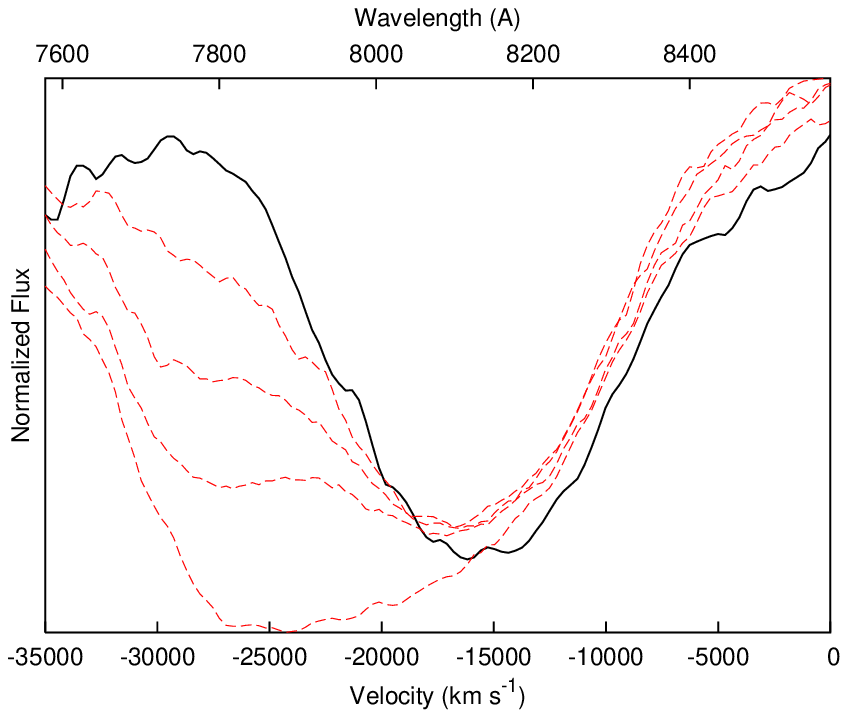} 
\end{tabular}
\caption{The geometry of Model B1 (left) and synthetic spectra obtained viewing 
this geometry from different orientations (right).
The opening angle of the blob is $\sim 80^{\circ}$. 
The red dashed lines in the right panel show the synthetic spectra corresponding to 
viewing angles of $0^{\circ}, 21^{\circ}, 33^{\circ}$ and
$42^{\circ}$, respectively, going from deep to shallow absorption. The angle is
measured from z-axis (the direction of the blob).
For angles closer to the equatorial plane the depth of the HVF 
is almost the same as in the spherical model without any enhancement (thick line).
\label{fig:blob1}}
\end{figure*}

First we describe the general behavior of HVFs in 3D models taking a model with
a large blob (Model B1) as an example. Figure \ref{fig:blob1} shows the geometry
(left)  and the synthetic line profile (right). In the left panel, the red
sphere shows the photosphere and the blue region shows the density enhancement.
The depth of HVFs in the synthetic spectra is different for different
lines-of-sight. Seen on the z-axis, which is defined as the direction of the
blob, the absorption at high velocity is deepest. If we move toward
the equatorial plane, the high velocity absorption becomes shallower and
shallower and it disappears when the line-of-sight reaches the edge of the
blob. Note that having the blob on one side has only a small effect on the
emission profile, because the volume occupied by the blob is small. The range of
profiles in the right panel of Figure \ref{fig:blob1} is qualitatively very
similar to Figure \ref{fig:SHV1D}, which shows that our 1D parametrization
captures the role of the blobs.If the blob is optically thick, the depth of
the HVFs is determined by the fraction of the projected photosphere 
that is concealed by the dense blob for each line-of-sight. 
We define as the ``covering factor'' and denote it as $f$.

Figure \ref{fig:blobSHV} shows that the HVFs obtained using different covering
factors in 3D models resemble those obtained with 1D synthetic spectra for
different values of $1-\exp (-\tau)$. The values are not exactly the same
because our 3D blobs are not infinitely optically thick. In fact, given a
typical optical depth $\tau_0$ in the blobs, one expects a relation $\exp
(-\tau) \sim (1- f) + f \exp (-\tau_0)$,  where $\tau$ is the line optical
depth introduced in the 1D computation. Thus $f = 1- \exp (-\tau)$ if $\tau_0 =
\infty$. Since $\tau_0$ has a finite value, $f \sim (1- \exp (-\tau))/(1 - \exp
(-\tau_0))$, which is smaller than the value in an infinitely thick case.
Figure \ref{fig:blobSHV} clearly indicates this. Comparing the value of $1-\exp
(-\tau)$ and $f$ that yield similar spectra shows that the effective $\tau_0$
of the 3D blob decreases with increasing angle. This is probably because the
average length of a segment crossing the blob is smaller for larger angles, as
there are more grazing trajectories.

If such a blob exists, the variation of the strength of the HVFs ($S_{\rm HV}$)
can be reproduced along various lines-of-sight without introducing unknown
parameters such as line optical depth $1- \exp(-\tau)$ in \S \ref{sec:HVF1D}.
We roughly classify the HVFs by their strength into three groups, strong HVFs
($f \gsim 0.7$), medium HVFs ($f \sim 0.45-0.7$) and weak HVFs ($f \lsim
0.45$). For example, SN 2002dj and SN 2003du have ``strong'' HVFs, SN 1999ee
has a ``medium'' HVF, and SN 2003cg is ``weak'' (see Fig \ref{fig:HVF}).

However, we have to consider the statistical properties of HVFs. Recent
observations suggest that a considerable fraction of the earliest spectra have
high velocity absorption. On the contrary, Model B1
suggests that strong HVF will be observed only $\sim 0.5$\% in all SNe. Figure
\ref{fig:histblob1} is a histogram of the covering factor $f$. It clearly
indicates that only very few orientations can produce strong HVFs and about
90\% of the observations ($f \lsim 0.45$) should show only the photospheric 
absorption.  Figure \ref{fig:histobs} shows the observed frequency of the HVF's strength made 
from Mazzali et al. (2005a and 2005b) by distributing 8 SNe into 4 bins.
The observed depth of the HVF is converted to the covering factor thorough
the numerical fit as in Figure 4.
It should be noted that the number of the sample is not enough in Figure 8
but the ubiquity of the HVFs is not overthrown 
even if we consider other SNe (Mazzali et al. 2005b).
The clear differences between the histogram of Model B1 
(Fig. \ref{fig:histblob1}) and the observation (Fig. \ref{fig:histobs}) 
are seen at $f \sim 0.0$ (too much in Model B1) and 
$\gsim 0.5$ (too small in Model B1). Although it is
of course not necessarily required that all observations are reproduced with a
single geometry, a general geometry may be inferred if a certain model can
reproduce the observed trend.  We investigate several geometrical models in the
following subsections and compare the distribution of synthetic properties to
the observed trend.

\subsection{Blob Models}

\begin{figure*}
\begin{tabular}{cc}
\includegraphics[scale=.80]{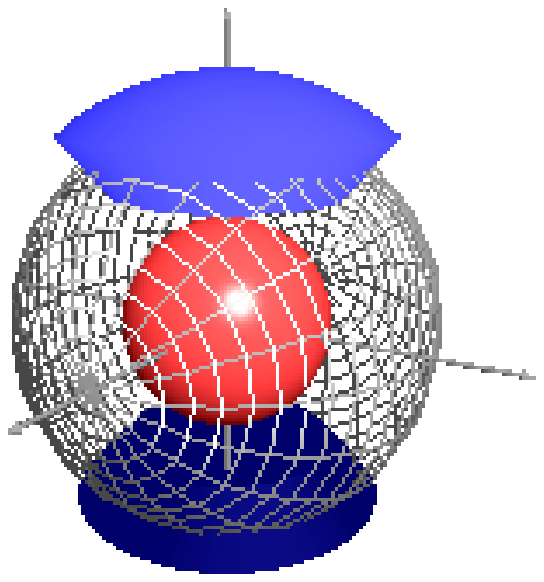} &
\includegraphics[scale=1.1]{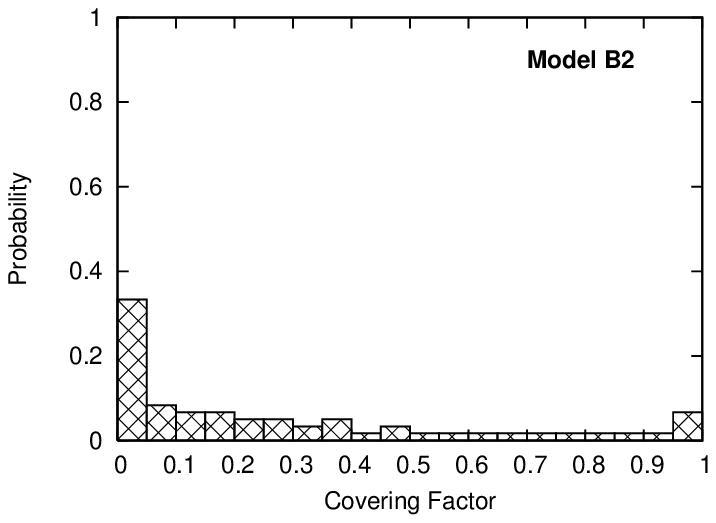} 
\end{tabular}
\caption{The geometry of Model B2 (left) 
and the distribution of the covering factor (right).
\label{fig:blob2}}
\end{figure*}

We showed that a model with one blob is unlikely as a general geometry of SNe
Ia. We therefore vary both the size of the blobs and their number. We show the
expected statistical properties of each model by means of the distribution of
covering factors, since we have shown that there is a direct correspondence
between $f$ and HVF strength.  Figure \ref{fig:blob2} shows the distribution of
covering factors in a model with two blobs (Model B2). As expected, the
fraction of lines of sight with large covering factor doubles and the fraction
of small factors decreases.  This is however still not consistent with the
fact that most spectra have high velocity absorption. 
Although we tested various blob sizes,
 most lines-of-sight have $f \lsim 0.4$ because of the large space
between the two blobs. We then increase the number of blobs adjusting their
size so that the photosphere is not completely covered.

\begin{figure*}
\begin{tabular}{cc}
\includegraphics[scale=.80]{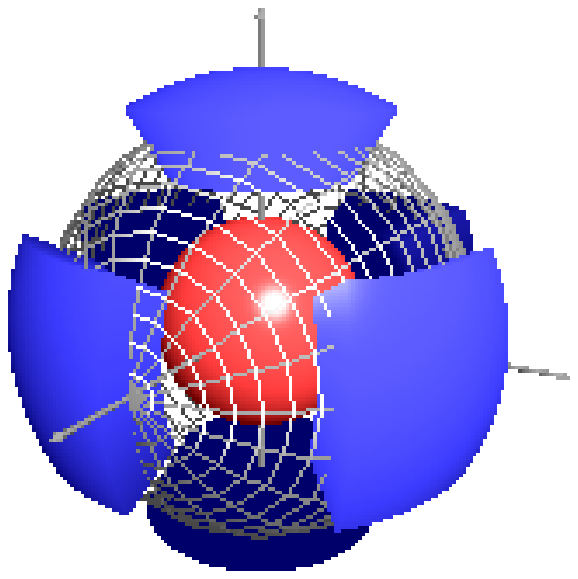} &
\includegraphics[scale=1.1]{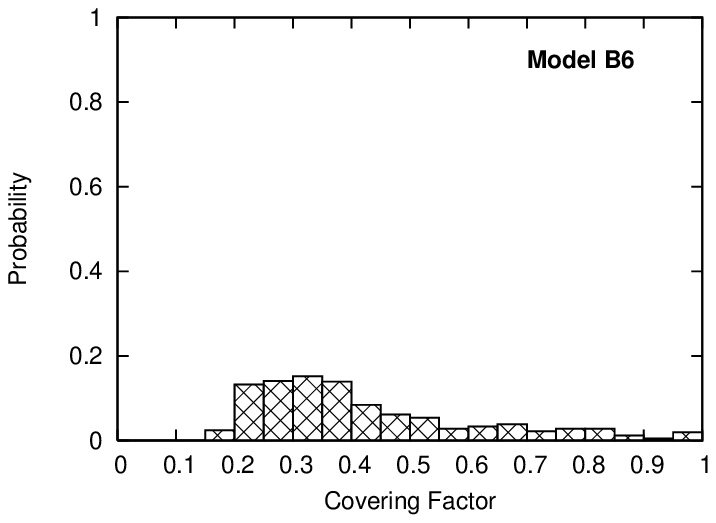} \\
\includegraphics[scale=.80]{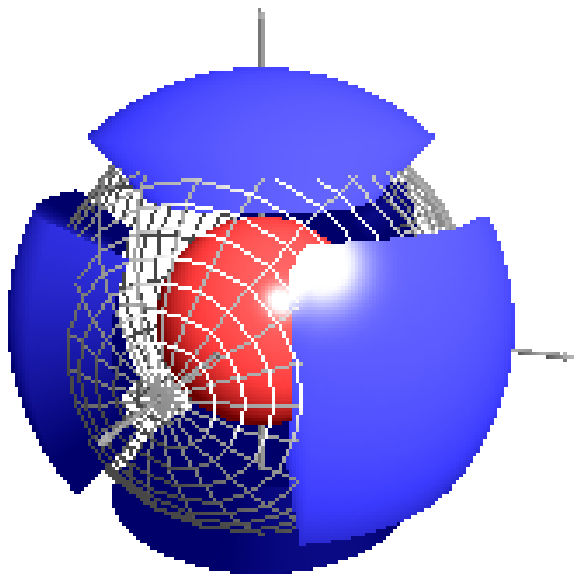} &
\includegraphics[scale=1.1]{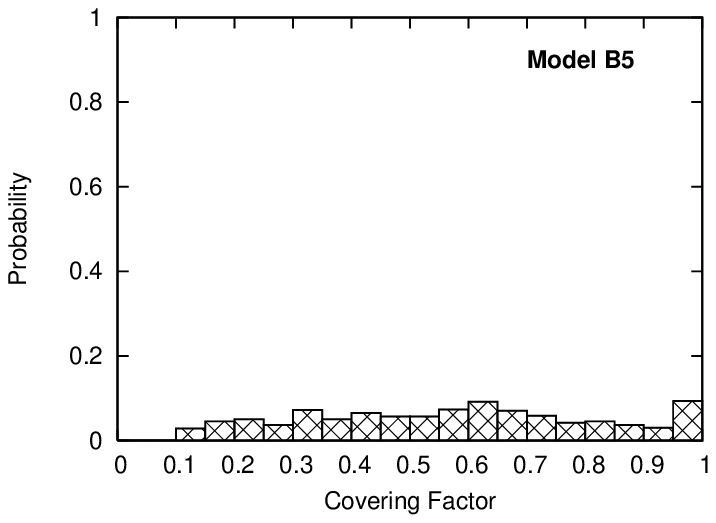} \\
\includegraphics[scale=.80]{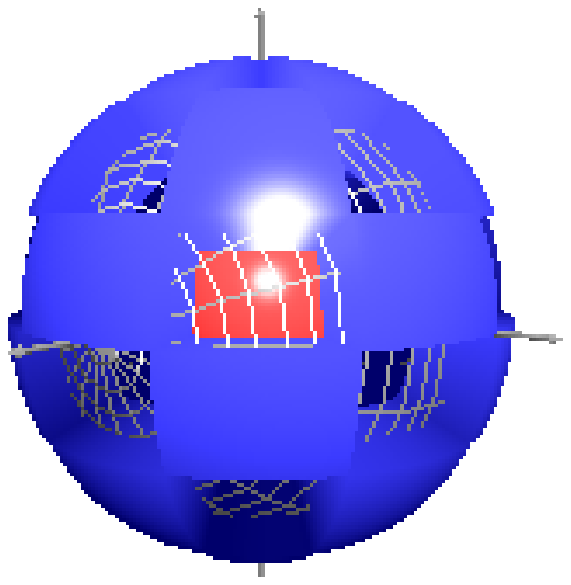} &
\includegraphics[scale=1.1]{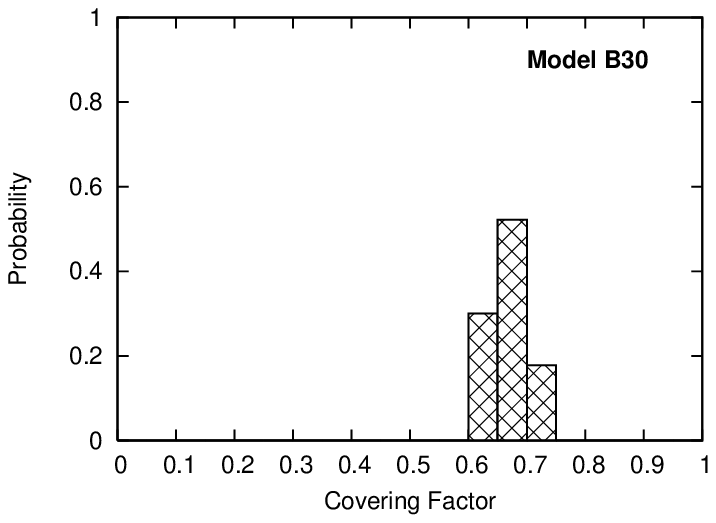} 
\end{tabular}
\caption{Top: the geometry of Model B6 (6 blobs with opening angle 60deg) and its 
distribution of covering factor.
Middle: same as Top but for Model B5 (5 blobs with opening angle 80deg). 
Bottom: same as Top but for Model B30 (30 blobs with opening angle 30deg) 
\label{fig:multi}}
\end{figure*}

Next we consider a case with several blobs. Figure \ref{fig:multi} shows
the distribution of covering factors for a model with 6 blobs with opening angle $60$
degrees (top panel; Model B6), a model with 5 blobs of $80$ degrees (middle
panel; Model B5) and a model with several blobs of $30$ degrees (bottom panel;
Model B30). From top to bottom, the space between the blobs becomes smaller.
Synthetic spectra show that the correlation between $f$ and the depth of the HVF
remains even when the blobs are crowded as in Model B30. Model B6 still has a
large space between the blobs, making the fraction of small $f$ large. In
addition, this model is very unlikely to produce the deep absorption required to reproduce
the objects like SN 2002dj and SN 2003du
(see, Fig \ref{fig:HVF} and Fig \ref{fig:HVFfit}),
which needs $f \sim 1$. 

Model B5 (Fig \ref{fig:multi}, middle) has smaller intervals between the blobs
and a large dispersion in the covering factor. This model has a 15\% probability
of showing a deep
absorption ($f \gsim 0.9$) while the fraction of lower $f$ is smaller than in any
of the models discussed before. The fact that HVFs exist in almost all early
spectra may suggest that a clumpy structure like Model B5 is a possible average
geometry.

If we cram more blobs (Model B30; Fig \ref{fig:multi}, bottom), the effect of
different lines-of-sight is averaged and the variation disappears. This result
is similar to that of the 1D test with $1 - \exp (-\tau) = 0.5$. Therefore we
can conclude that this geometry is not typical of the majority of SNe Ia,
although it may exist.

\subsection{Torus Models}

\begin{figure*}
\begin{tabular}{cc}
\includegraphics[scale=.80]{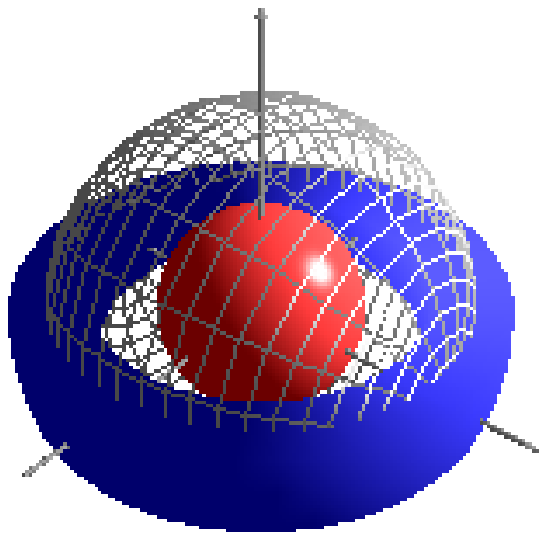} &
\includegraphics[scale=1.1]{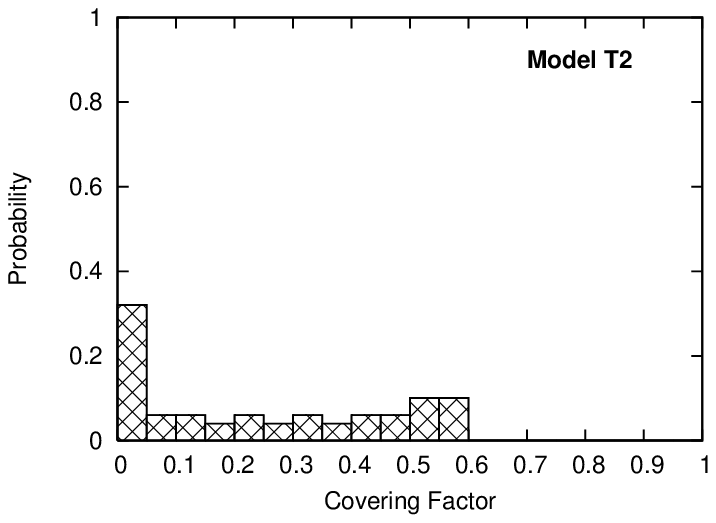} \\
\includegraphics[scale=.80]{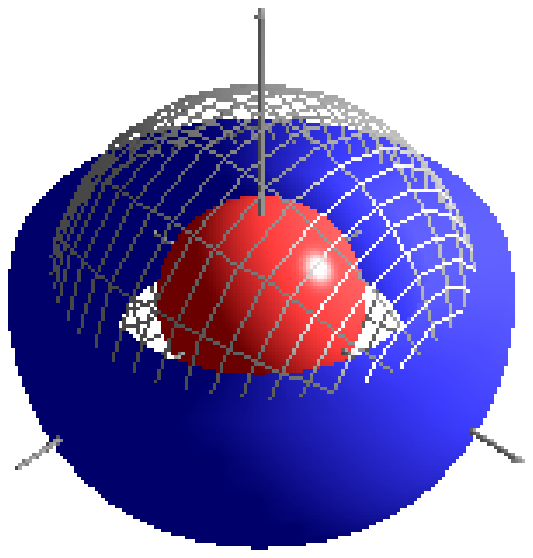} &
\includegraphics[scale=1.1]{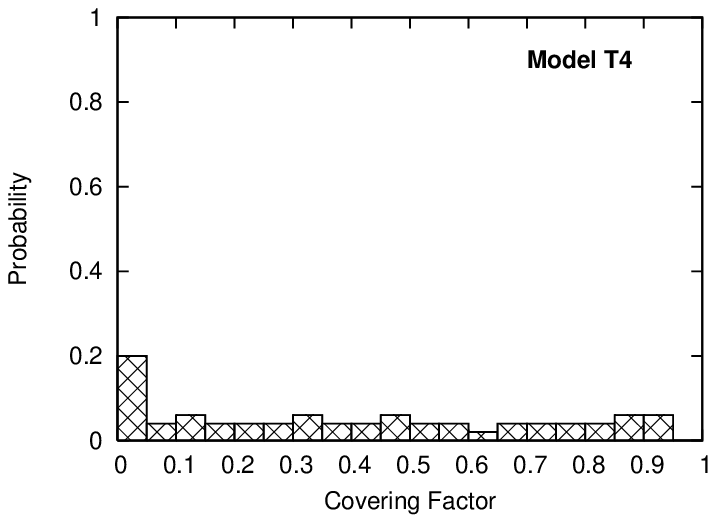} \\
\includegraphics[scale=.80]{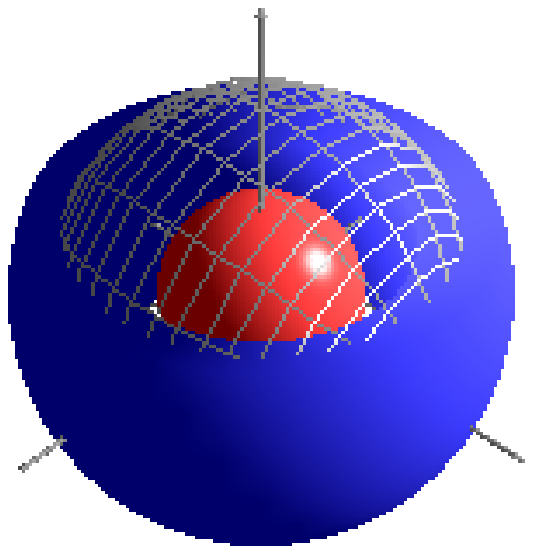} &
\includegraphics[scale=1.1]{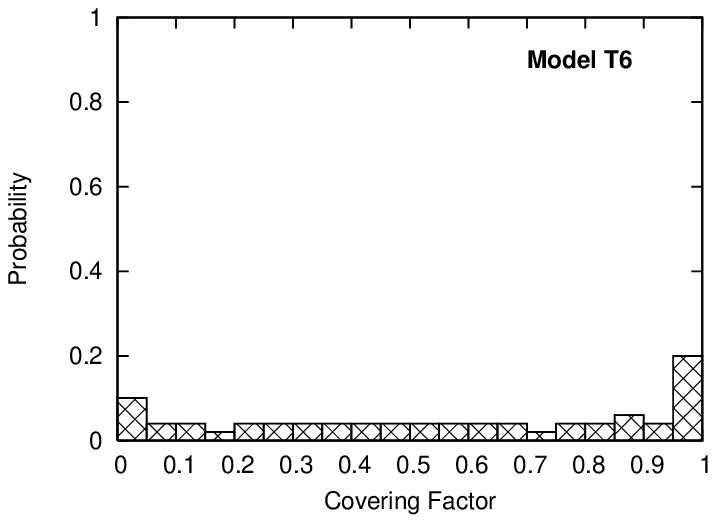} 
\end{tabular}
\caption{Same as Figure \ref{fig:blob2} but for torus Model T2 (opening angle 20 deg, 
top), Model T4 (opening angle 40 deg, middle) and Model T6 (opening angle 60 deg, 
bottom).
\label{fig:torus}}
\end{figure*}

We distinguish a model that has a torus-like density enhancement from blob
models because the origin of the enhancement is expected to be interaction with
CSM or an accretion disk in this case. The main problem with the hypothesis
that HVFs come from CSM interaction is the fact that extremely high mass loss
rates are required to place the material just outside the explosion. One
expects that this may be solved if the progenitor WD is surrounded by an
accretion disk and the ejecta collide with the disk, leading to density
enhancements on the plane of the disk. Figure \ref{fig:torus} shows the
distribution of covering factors of models with torus-like enhancements (Model
T2, T4 and T6). These models have a disk-like enhanced region with opening
angles of $20^{\circ}, 40^{\circ}$ and $60^{\circ}$, respectively. As shown in 
the left panels of Figure \ref{fig:torus}, the $60^{\circ}$ disk has a 
thickness comparable to the diameter of the photosphere. 

In Model T2 the disk is too thin for the covering factor ever to become large
(top panels in Fig. \ref{fig:torus}). If the thickness increases to
$40^{\circ}$ (Model T4; middle panels in Fig. \ref{fig:torus}), the fraction of
``weak'' HVFs decreases while those of ``strong'' and ``middle'' HVFs increase.
If the the disk is thick enough for the photosphere to be totally concealed,
the fraction of ``strong'' HVFs increases (Model T6; bottom panels in Fig.
\ref{fig:torus}) and the distribution of covering factors becomes broad. Only
Model T6 may produce the observed distribution of HVF strength. 

This implies that a very thick disk is required, with thickness comparable to
the diameter of the WD. The material in the disk should be swept by the SN ejecta
in a radial direction leaving an imprint of the disk's angular size
after the explosion. While this may be a realistic model, the absence of
hydrogen lines in the early spectra remains the strongest argument 
against the disk origin of the HVFs.  
This problem can be overcome by assuming low temperature ($T \sim 4,500K$) 
at the shocked ejecta (Gerardy et al. 2004),
where the hydrogen lines are less active than metal lines. 
And see the consideration made in Mazzali et al. (2005a) 
on the role of hydrogen in making the CaII IR triplet broad. They
estimate M(H)$\sim 0.004 \Msun$ in spherical symmetry.

\section{Discussion}
\label{sec:dis}

The strength of the HVFs ($S_{\rm HV}$) was parameterized via the line optical
depth of the CaII IR triplet in \S \ref{sec:HVF1D}. The observed range of HVFs
can be reproduced using only geometrical effects if we assume that the
optically thick region distributes discretely. The covering factor that
represents how the photosphere is concealed by the optically thick region acts
exactly as $1 - \exp (-\tau)$ in \S \ref{sec:HVF1D}. Among the various
geometries discussed in the previous section, Models B1 and B2 have a very
large fraction of lines-of-sight with $f \lsim 0.4$, which makes almost no
HVFs. Since observations suggest that a considerable fraction of the earliest
spectra show HVFs, these models can be ruled out as a standard configuration.

As the interval between the blobs becomes smaller, the fraction of
lines-of-sight with $f \lsim 0.4$ decreases (\eg from Model B6 to Model B30).
However, too crowded structures like Model B30 do not give rise to a variation
in the covering factor and so the strength of HVF is constant. Model B5 has a
wide range of covering factors from $f \sim 0.1$  to $f \sim 1$. If such a
structure is formed from the explosion, it can explain the observed variety
naturally.

It should be noted that models with a hole can be tuned to give the observed
distribution of HVF more easily than models with ``blobs''  
by changing the covering factor, \ie the size of the hole.
How such geometries might be produced is
however unclear, as they still require a dense high velocity region and a line
of sight that is never far from the hole, because otherwise most SNe would show
strong HVFs. 

A torus model may reproduce the observed distribution of HVF if the disk is
geometrically thick ($\sim 60^{\circ}$).
Such a thick disk-like enhancement could be produced 
if the accretion disk is thick enough to surround the WD before the explosion.
The opening angle of the accretion disk is likely to be reflected  
on the angular size of a density enhancement.

  A torus in Model T2 or T4 is so thin that it provides a large fraction of 
``weak'' features, which are not frequently observed.
However, it may be possible to reduce the fraction of $f \lsim 0.45$ 
if the interaction with CSM resulting from spherical or bipolar WD wind
occurs.
Although many additional parameters should be introduced 
to investigate the possibility of this scenario, 
this scenario seems to work qualitatively.

When the optically thick region at high velocity is introduced as a density
enhancement, other features are also affected. The SiII $\lambda$6355 line is
the most notable line, and its behavior has been thoroughly examined (\eg
Benetti et al. 2005).  Figure \ref{fig:HVFSi} shows spectra around SiII
$\lambda$6355. If the photosphere is totally covered by dense blobs, the
absorption minimum becomes $v \sim 22,500$\kms.
Such high velocity features have not been observed. Also,
the velocity of SiII $\lambda$6355 is not always 
correlated with that of the CaII IR triplet in the observed spectra.
For example, SN 2002bo has a higher SiII $\lambda$6355 velocity 
$\sim 15,500$\kms (at -8 days after the maximum) 
than that of SN SN 2003kf $\sim 12,000$\kms (at -9 days)
but the CaII HVF of SN 2002bo ($v \sim 22,000$\kms) is slower than
that of SN 2003kf ($v \sim 23,500$\kms).
Therefore the assumption that HVFs in the CaII IR
triplet are due only to density enhancements may not be correct. However there
are some SNe that have an extended blue wing in SiII $\lambda$6355. This may
indicate some degree of density enhancement also occurs. The combination of
density enhancement and abundance enhancement may provide a proper correlation.
Such a study, however, is beyond the scope of this work. It is worth noting that the mass
of additional material is reduced from $\sim 0.1 \Msun$ in the spherical case
to $\sim 0.05 \Msun$ in Models B5, B6 and B30.

The velocity range of HVFs ($v_{\rm HV}$) is found to be the same 
in all the models and all the lines-of-sight
as long as the position of the enhancements is fixed.
If the covering factor is identical, 
one might expect that the position of the absorption minimum 
depends on the line-of-sight.
However, we found that this effect is very slight.
The strength of the photospheric absorption $S_{\rm ph}$ is
not correlated with the covering factor 
and it is not affected even if the line-of-sight is varied.

\section{Conclusions}
\label{sec:con}

We tested the properties of HVFs of CaII IR triplet through 1D and 3D
modeling. Parameterized 1D simulations were used to extract the main features
that determine the HVF. We defined three parameters that govern the HVFs. These
are the strength of the HVF $S_{\rm HV}$, its velocity range $v_{\rm HV}$, and
the strength of the photospheric component $S_{\rm ph}$. As for $S_{\rm HV}$,
it was shown that geometrical effects on models based on high velocity blobs or
a thick torus can provide enough variety to cover all the observations. On the
contrary, $v_{\rm HV}$ is not affected by line-of-sight effects. Therefore we
speculate there may be a number of blobs that have different velocities. This
might be possible if the WD rotates and the clumpy structure formed by the
deflagration flame has different properties depending on direction. 
Such a variety is not produced in a torus model.
However, it may be possible that different disks lead to different 
enhancements from SN to SN
since the degree of density enhancement and its velocity range 
depend strongly on the density, 
the total mass and radial extent of the disk.

The strength of the photospheric absorption ($S_{\rm ph}$) may be influenced by
the temperature and the abundance. Although we changed this value by changing
the abundance in this paper, it should be studied  whether there is any
correlation between the temperature and $S_{\rm ph}$. This requires knowing the
exact temperature and  photospheric velocity by fitting observed spectra, 
which will be the subject of future work (M. Tanaka et al. in preparation).

The models that best explain the statistics of the observed strength of the HVFs
are a blob model (B5) with $\sim 5$ blobs of $80^{\circ}$ and a thick torus of
opening angle $\sim 60^{\circ}$.  
While a torus model sounds appealing, the main argument against it 
is the absence of hydrogen lines 
(but see Gerardy et al. 2004, and Mazzali et al. 2005a) and 
the variation of the HVF velocity.
A blob model may result from the explosion if the mushroom
structure of the deflagration is not completely washed away in a delayed
detonation (Gamezo et al. 2005).  Although this does not mean that all the SNe
have a single geometry, the average structure may have a distribution of high
velocity material similar to that of Model B5. 
To verify this conclusion, more early phase 
observations before maximum are needed. 
A statistical study can constrain the structure more accurately. 

Model B5 is different from the 3D structure that was obtained for SN 2001el by
Kasen et al. (2003). Their model has one large blob and an aspherical
photosphere, and is able to reproduce the observed polarization flux. Our
results suggest that such a geometry cannot be standard (see Model B1 in \S
\ref{sec:results}). It is completely unknown whether only polarized SNe Ia have
such a geometry, consisting of one or two blobs in the outermost region. 
Rotation of the WD may lead to such a configuration. Although a
statistical study of polarization is not available yet, 
it is potentially useful to
verify the geometry, including the abundance distribution.
In addition, the statistical study of polarization may be able to 
distinguish the torus models and the blob models because the torus-like
enhancement tend to produce a large polarization (Kasen et al. 2003).

Detailed hydrodynamic calculations of 3D deflagration flames 
suffer from shortage of high velocity material. 
It is unknown whether such a result indicates our
ignorance about the deflagration flame or the existence of a transition from a
deflagration to a detonation. 
It is therefore quite important to define a suitable
explosion mechanism from the observational data. 
This will be achieved via a
statistical study presented in this paper 
when larger number of early spectra are available. 
The importance of early observations of SNe Ia 
cannot be understated.

\acknowledgments
This work has been supported in part by the Grant-in-Aid for
Scientific Research (16540229, 17030005, 17033002) and the 21st
Century COE Program (QUEST) from the JSPS and MEXT of Japan.
K.M is supported through the JSPS 
(Japan Society for the Promotion of Science) 
Research Fellowship for Young Scientists.

\clearpage

\plotone{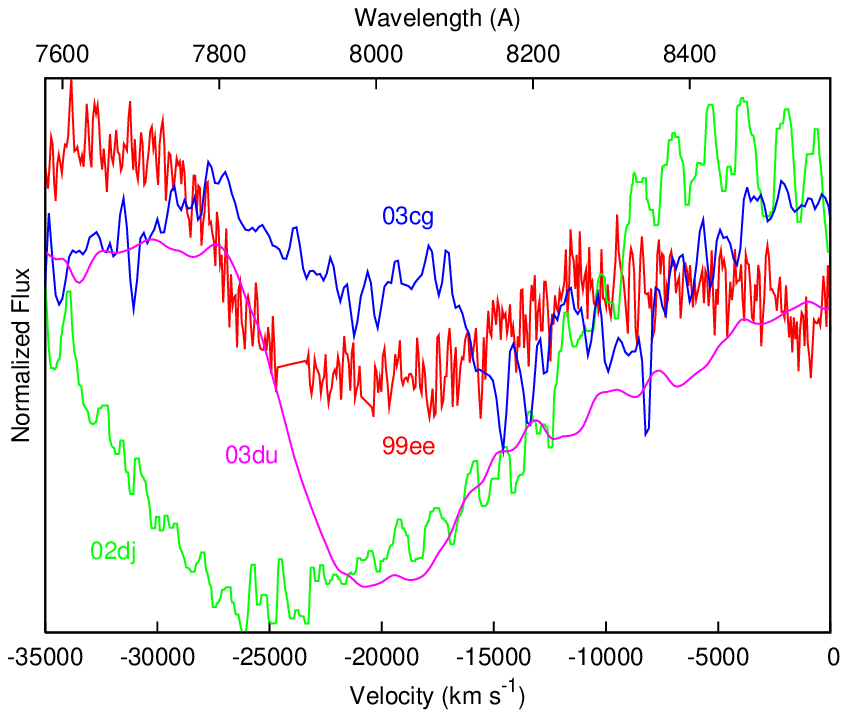} 
\figcaption{Line profiles of the CaII IR triplet in the early spectra of 4 SNe.
Shown are SN 1999ee (9 days before maximum, Hamuy et al. 2002, red); SN 2002dj 
11 days before maximum, Kotak et al. 2005, green);  SN 2003cg (8 days before 
maximum, Elias-Rosa et al., in preparation, blue); and SN 2003du (11 days 
before maximum, Stanishev et al., in preparation, pink).  
The flux was adjusted to have the same continuum flux at $\lambda \sim 
7500 - 8500$\AA\ by multiplying by 
a constant. 
While the HVFs in SN 2002dj and SN 2003du are deep, that of SN 1999ee is weak.
SN 2003cg has almost no high velocity absorption.
\label{fig:HVF}}

\plotone{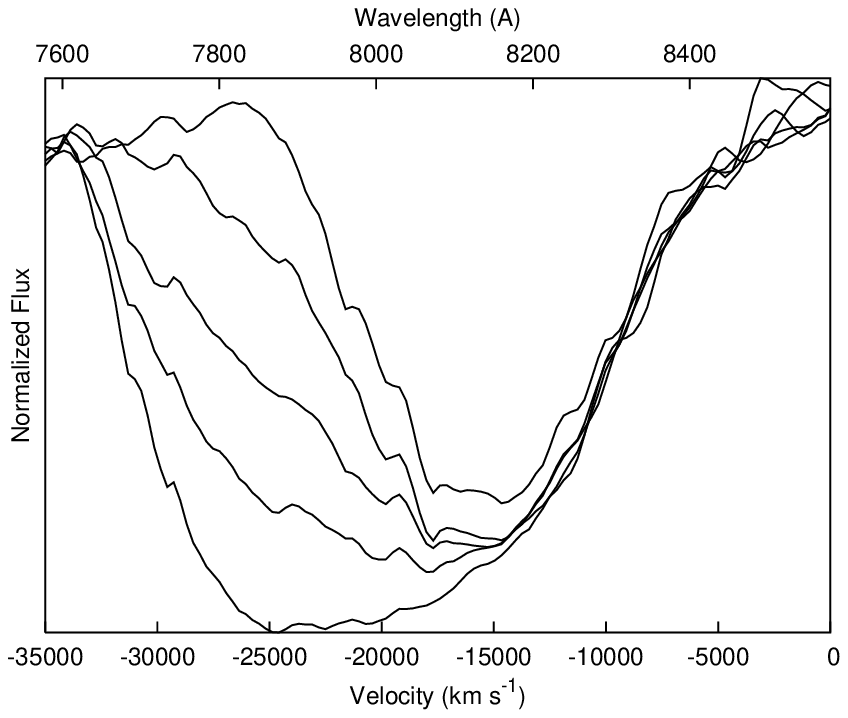} 
\figcaption{The effect of a variation in the strength of high velocity absorption 
$S_{\rm HV}$ on the CaII IR triplet profile. The line optical depth is
parametrized as $1-\exp (-\tau)$. The various profiles, going from 
deeper to shallower absorption, have $1-\exp (-\tau) = 0.9, 0.7, 0.5, 0.3$ and
$0.1$, respectively, at $v \sim 22,000 - 29,000$\kms. A variation of
these parameters does not affect the photospheric absorption at all.
\label{fig:SHV1D}}

\plotone{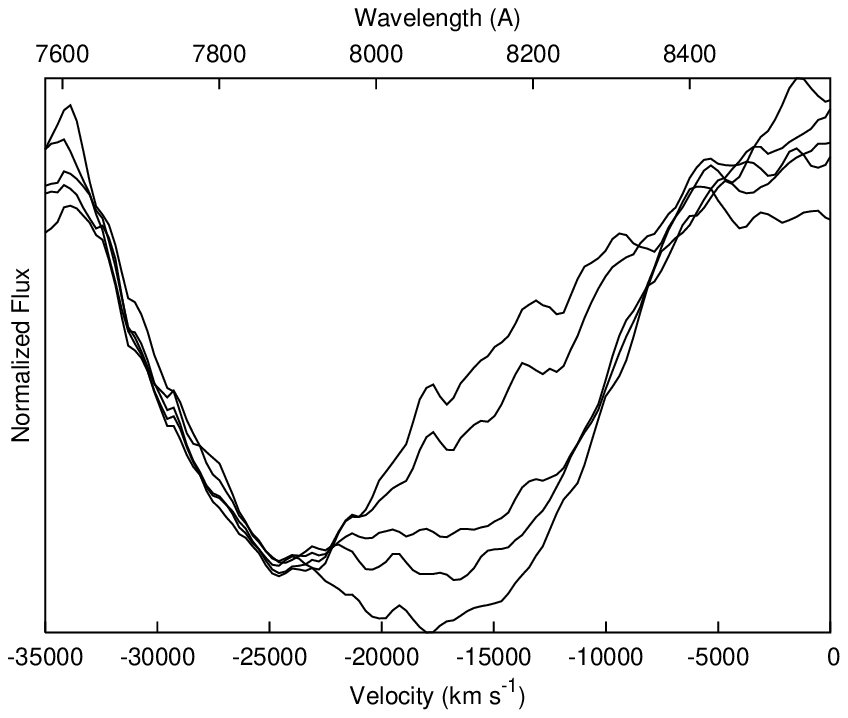}
\figcaption{The effect of a variation in the strength of photospheric absorption
$S_{\rm ph}$ on the CaII IR triplet profile.
The strength is changed through the abundance of Ca. The various profiles, going 
from deeper to shallower, correspond to a mass fraction $X({\rm Ca}) = 0.068, 0.034, 0.017,
0.008$ and $0.004$, respectively.
The high velocity absorption is not affected at all because we fix the 
value of $1 - \exp (-\tau) = 0.7$ and the velocity separation between the
photospheric component and the HVF is too large. 
\label{fig:Sph1D}}

\clearpage
 
\clearpage
\plotone{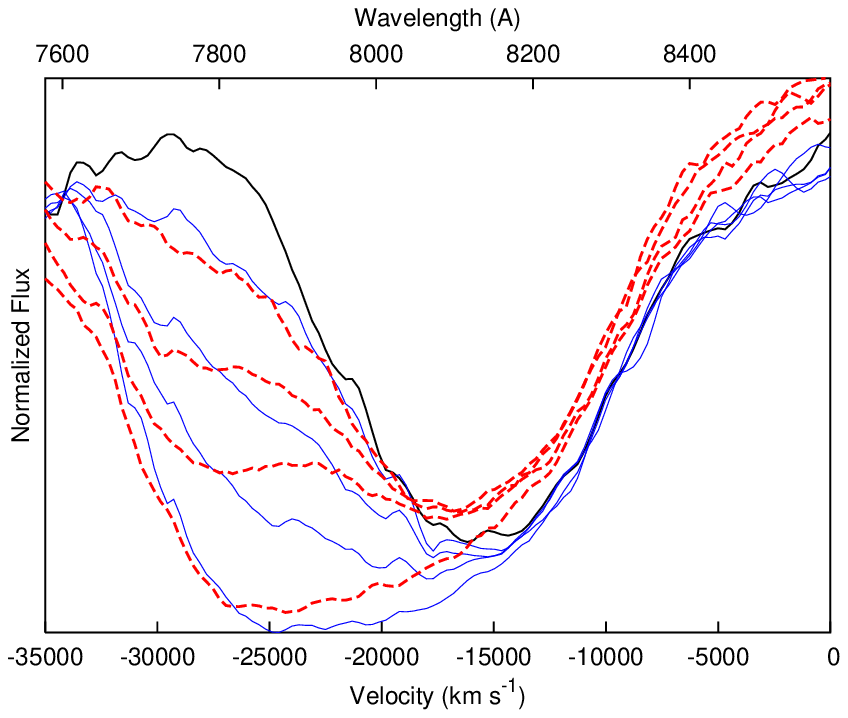} 
\figcaption{3D synthetic spectra computed from Model B1 (red) compared with the 1D
results of Figure 2 (blue). While the 1D results have been obtained for shells with
different opacities, the 3D spectra correspond to the same model viewed from
different angles, and therefore differ by the covering factor $f$. The black line
is the model without high velocity enhancement, or $f = 0$ in 3D. 
The 1D spectra have $1-\exp (-\tau) = 0.9, 0.7, 0.5$ and
$0.3$, from deeper to shallower.
The 3D spectra have covering factor $f =  1.0, 0.88, 0.66$ and
$0.46$, from deeper to shallower. 1D and 3D results are related.
The difference between the values of $f$ and $1-\exp (-\tau)$ comes from
the fact that the blob in the 3D model is not infinitely thick 
(see \S \ref{sec:results}).
\label{fig:blobSHV}}

\plotone{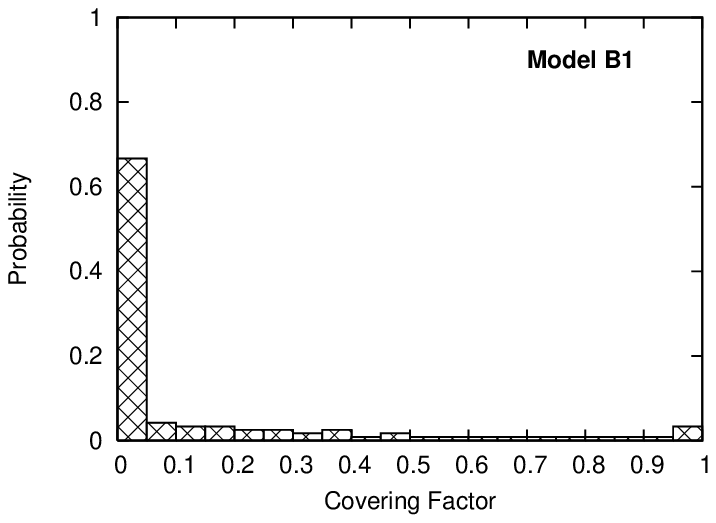} 
\figcaption{Probability distribution of the covering factor $f$ in Model B1.
The horizontal axis shows the covering factor and
the vertical axis shows the frequency of the covering factor 
if we see this model from various lines-of-sight.
\label{fig:histblob1}}

\plotone{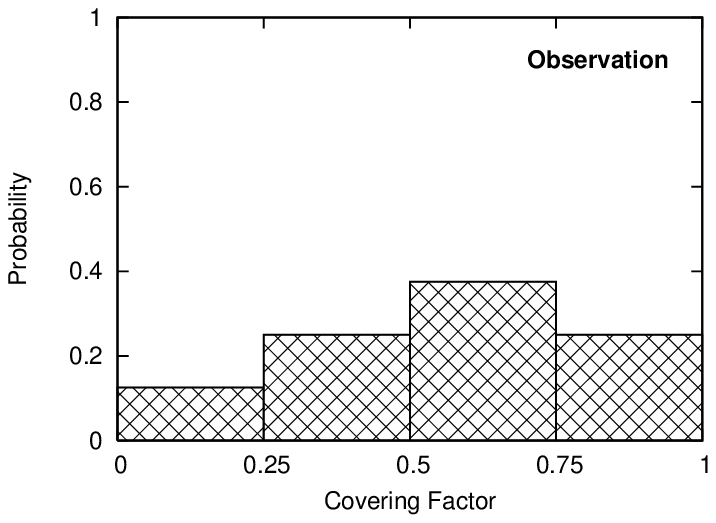} 
\figcaption{Probability distribution of the observed HVFs.
8 SNe from Mazzali et al. (2005 and 2005b) are distributed to 4 bins by
fitting the depth of the HVFs.
Note the number of the bin is not identical with the other histogram of
the covering factor $f$.
\label{fig:histobs}}

\clearpage

\clearpage

\clearpage

\clearpage
\plotone{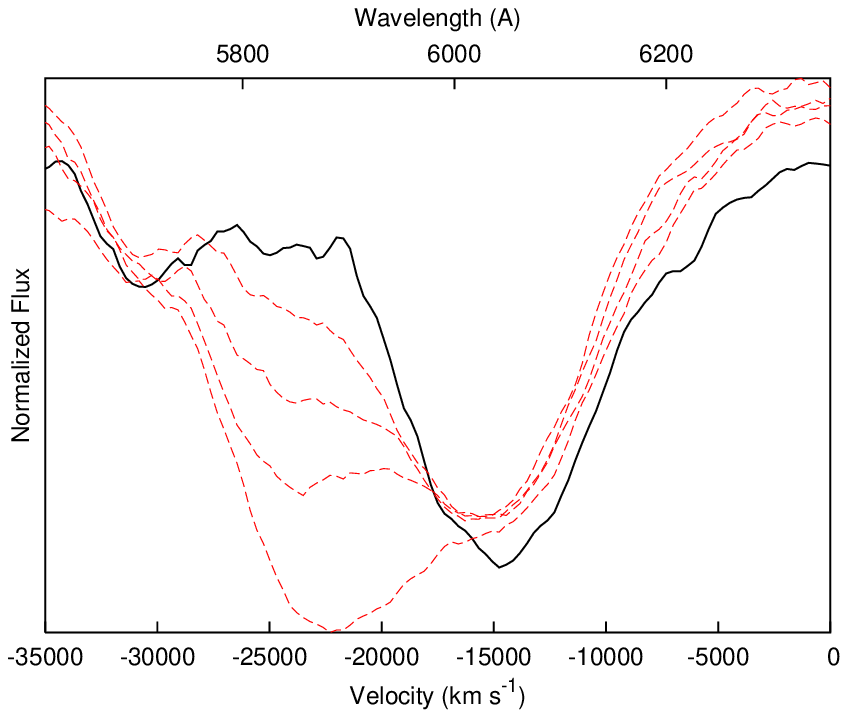} 
\figcaption{Synthetic spectra around SiII $\lambda$6355.
The thick line is a 1D spectrum without any enhancement.
Dashed lines are 3D spectra with covering factor $f = 1.0, 0.88, 0.66$ 
and $0.46$ from deeper to shallower.
If the photosphere is totally hidden ($f \sim 1$), 
the feature forms at a high velocity that has never been observed in any SN.
\label{fig:HVFSi}}

\end{document}